\documentstyle[11pt,paspconf,epsf]{article}

\begin{document}

\title{Gamma-ray Pulsars in a Modified Polar Cap Scenario}
\author{B. Rudak and J. Dyks}
\affil{N. Copernicus Astronomical Center, Rabia\'nska 8, 87-100 Toru\'n,
Poland}

\begin{abstract}
We present a polar-cap model which incorporates a likely acceleration of
Sturrock pairs with their subsequent
contribution to gamma-ray luminosity $L_\gamma$. This model reproduces 
$L_\gamma$ for seven pulsars detected with {\it Compton Gamma Ray Observatory} 
experiments, avoiding at the same time the problem
of the empirical gamma-ray
death line of Arons (1996).

Also, we estimate 
the efficiency of reversing newly created positrons by residual
longitudinal electric field. Over the wide range of spin-down
luminosity values the predicted polar-cap X-ray luminosity $L_{\rm X}({\rm pc})$
goes as $\propto L_{\rm sd}^{0.6}$.				
Model calculations for B0823+26, B0950+08, B1929+10,
and J0437-4715 are compared with existing observational constraints
on thermal X-ray components.
\end{abstract}


\section{Introduction}
We present a semi-empirical polar cap model
which reproduces the gamma-ray luminosities 
inferred for seven pulsars (we dub them {\it Seven Samurai})
detected with the CGRO experiments: Crab, B1509-58, Vela, B1951+32, B1706-44,
Geminga, and B1055-52.
The model avoids the problem of the empirical gamma-ray
death line of Arons (1996). We also address
the question of expected
thermal X-ray emission due to polar cap reheating by reveresed particles. With numerical
calculations used to obtain $L_\gamma$ we tried to reproduce recent {\it ROSAT} 
and {\it ASCA} results.

\section{Gamma-ray Luminosities}
The model (see Rudak \& Dyks 1997 for details) is
based on polar cap activity triggered by ultrarelativistic
primary electrons as described by Daugherty \& Harding (1982).
Our two basic assumptions are:\hfill\break
1) The energy of primary electrons $E_0$ is only a few times higher than the threshold energy
required to induce pair creation in a dipolar magnetic field
of surface strength $B_{12}$ Teragauss, with other
restrictions applied when necessary:    
\begin{equation}
E_0 = \min \{\zeta \cdot E_{\rm min}, \,E_{\rm W}, \,E_{\rm max}\}, 
\end{equation}
where $\zeta$ satisfies $2 \la \zeta \la 5$, and  $E_{\rm min} = 1.2 \times 10^7 \, B_{12}^{-1/3} P^{1/3}$,
$E_{\rm W} = 1.2 \times 10^7 \, B_{12} P^{-2}$, $E_{\rm max} = 4.6\times 10^7 \, B_{12}^{1/4} P^{-1/8}$
(Rudak \& Ritter 1995). \hfill\break 
2) Out of $n_{\pm}(E_0)$ Sturrock pairs produced per each primary electron a fraction
$\eta_\gamma$ is subject to further acceleration and contributes to pulsar's gamma-ray emission. 

We obtain thus a following prescription for $L_\gamma$:
\begin{equation}
L_\gamma = \eta_\gamma \cdot E_{\pm} \cdot n_{\pm} \cdot \dot N_{\rm GJ}, 
\end{equation}
where $E_{\pm}$ -- a characteristic energy attained by secondary particles due to acceleration -- is
assumed to be of the order of $E_0$; 
the parameter $\eta_\gamma$ will be determined by comparing results of numerical simulations
of creation of Sturrock pairs with the {\it CGRO} results;
$\dot N_{\rm GJ}$ is the Goldreich-Julian rate of outflow of primary electrons.

\begin{figure}
\begin{center}
\leavevmode
\epsfxsize=8.4 cm 
\epsfbox{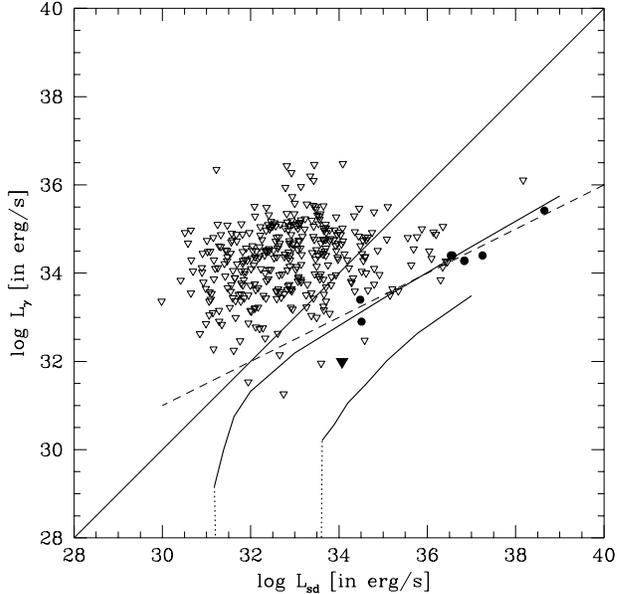}
\end{center}
\caption{Total $L_\gamma$ inferred from {\it EGRET, COMPTEL}, and {\it OSSE}
is shown for seven pulsars (filled dots).
Open triangles are the {\it EGRET} upper limits (Nel et al.1996).
Filled triangle marks the {\it EGRET} U.L. for the millisecond pulsar J0437-4715.
The dashed line represents 
$L_\gamma = 10^{16} \cdot L_{\rm sd}^{1/2}$. The upper solid line 
is the evolutionary track calculated for $B = 10^{12}$G, the lower one is for
$10^9$G.
}
\end{figure}
 
The model was confronted with the gamma-ray luminosities for seven pulsars inferred from
available data from {\it OSSE} (Schroeder et al.1994), {\it COMPTEL} 
(Carrami{\~n}ana et al.1995), and {\it EGRET} (Fierro 1995, Nel et al.1996) experiments
(assuming gamma-ray beaming angle $\Omega_\gamma = 1 {\rm sr}$ in all cases). 
Two evolutionary tracks in the $L_\gamma - L_{\rm sd}$ space calculated for
a classical pulsar ($B = 10^{12}$G), and for
a millisecond pulsar ($B = 10^9$G) are shown in Fig.1.

For a fixed value of
$L_{\rm sd}$, the predicted $L_\gamma$ depends rather weakly on magnetic field strength $B$ as long as
$10^{11}{\rm G} \la B \la 10^{13}{\rm G}$, in contrast to other existing models. 
Due to this property the model
reproduces the seven detections equally well as e.g. a simple empirical relation   
$L_\gamma \propto L_{\rm sd}^{1/2}$ does. 
At the domain of millisecond pulsars
($\sim 10^8 - 10^9{\rm G}$), $L_\gamma$~ drops significantly. 
The value of parameter $\eta_\gamma$ in Eq.2 is $\sim 4\times 10^{-3}$.

The upper limit for J0437-4715 from {\it EGRET} is about one order of magnitude
above the expected level. Out of four classical pulsars with {\it EGRET} upper limits
visibly below our predictions (see Fig.1) B1046-58 and B1929+10 
pose no threat to the model
due to existing upper limits set by {\it COMPTEL}. There is no detailed information
available about the other two - B0950+08 and B0656+14.

\section{Reheating of the Polar Cap}
Some positrons, before reaching a region of further acceleration, may
be stopped by local residual longitudinal electric field
and directed back to the surface.

\begin{figure}
\begin{center}
\leavevmode
\epsfxsize=8.4cm \epsfbox{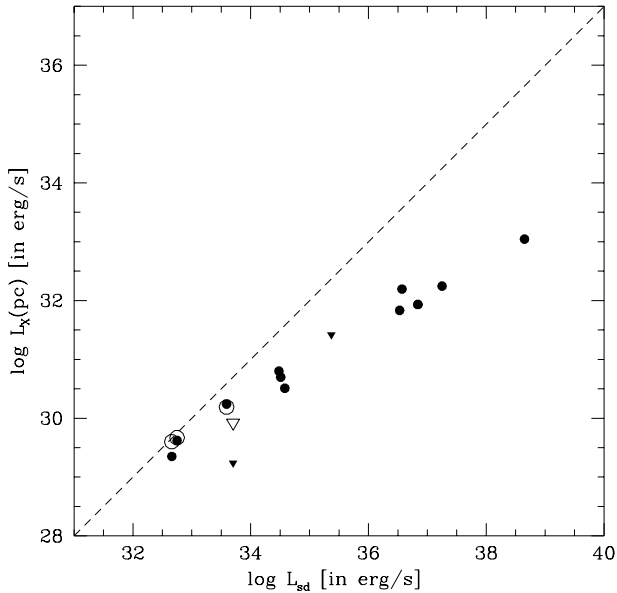}
\end{center}
\caption{Thermal X-ray components calculated for 
{\it Seven Samurai} along with 4 other classical pulsars (filled dots), 
and 2 millisecond pulsars (filled triangles). The open triangle is
the upper limit for J0437-4715 (Halpern et al.1997).
Open circles are possible thermal components for B0823+26 (Becker 1995),
B0950+08 and B1929+10 (Wang \& Halpern 1997).
}
\end{figure}

Each returning positron acquires on its way
an energy comparable to the energy of the primary electron,
$E_+ \simeq E_0$, and then deposits it onto the polar cap surface.
A heat conductivity inwards
the star, and across magnetic lines is negligible and 
dissipated energy is radiated out on the spot,
predominantly in form of thermal X-rays, with the luminosity $L_{\rm X}({\rm pc})$.

If out of all positrons ($\dot N_+ = n_\pm \cdot \dot N_{\rm GJ}$) a fraction $\eta_\downarrow$
will be reversed, then
\begin{equation}
L_{\rm X}({\rm pc}) = \eta_\downarrow \cdot E_0 \cdot n_\pm \cdot \dot N_{\rm GJ}. 
\end{equation}
Suppose now that $\eta_\downarrow \propto \epsilon_\parallel^{-1}$, where
$\epsilon_\parallel$ is a mean energy of freshly created pairs retained
in their longitudinal motion (Rudak 1997, in preparation).

Fig.2 shows $L_{\rm X}({\rm pc})$ as a function of spin-down luminosity 
$L_{\rm sd}$ calculated for 
{\it Seven Samurai}, 4 other classical pulsars, and 2 millisecond pulsars. 
Over the wide range of $L_{\rm sd}$ values the calculated polar-cap X-ray luminosity behaves as 
$L_{\rm X}({\rm pc}) \propto L_{\rm sd}^{0.6}$. This formula is significantly flatter
than the empirical relation
$L_{\rm X} \approx 0.001 L_{\rm sd}$ of Becker \& Tr{\"u}mper (1997).
Even though some models are able to reproduce this relation with non-thermal X-ray spectra
quite satisfactorily
(Cheng et al.1997), it seems very likely that actual X-ray spectra consist of two components:
a non-thermal one - due to synchrotron radiation, and a thermal one - due to polar-cap reheating.
For high $L_{\rm sd}$ pulsars the former component dominates, while at the
low-$L_{\rm sd}$ end the latter component becomes visible.

The values of parameter $\eta_\downarrow$ from Eq.3 are on average
$\sim 10^{-5}$ and $\sim 10^{-3}$ for classical and millisecond pulsars, respectively.


\acknowledgments
This research was financed by the KBN grant 2P03D 00911,
and the Jumelage program
``Astronomie Pologne". BR is grateful to
Jean Pierre Lasota for his warm hospitality in DARC (Meudon).

\end{document}